\documentclass{midl}

\usepackage{siunitx}
\usepackage{url}

\newcommand{\netname}{SynthSeg}

\begin{document}


\jmlrworkshop{Medical Imaging with Deep Learning (MIDL) 2020}

\title[A Learning Strategy for Contrast-agnostic MRI Segmentation]{A Learning Strategy for Contrast-agnostic \\ MRI Segmentation}

\midlauthor{%
\Name{Benjamin Billot\nametag{$^{1}$}} 
\Email{benjamin.billot.18@ucl.ac.uk}\\
\Name{Douglas N. Greve\nametag{$^{2}$}}
\Email{dgreve@mgh.harvard.edu}\\
\Name{Koen Van Leemput\nametag{$^{2,3}$}}
\Email{kvle@dtu.dk}\\
\Name{Bruce Fischl\nametag{$^{2,4,5}$}}
\Email{fischl@nmr.mgh.harvard.edu}\\
\Name{Juan Eugenio Iglesias\midljointauthortext{Contributed equally}\nametag{$^{1,2,4}$}} \Email{e.iglesias@ucl.ac.uk}\\
\Name{Adrian V. Dalca\midlotherjointauthor\nametag{$^{2,4}$}} 
\Email{adalca@mit.edu}\\
\addr $^{1}$ Center for Medical Image Computing, University College London, UK \\
\addr $^{2}$ Martinos Center for Biomedical Imaging, Massachusetts General Hospital and Harvard Medical School, USA\\ 
\addr $^{3}$ Department of Health Technology, Technical University of Denmark, Denmark\\ 
\addr $^{4}$ Computer Science and Artificial Intelligence Laboratory, Massachusetts Institute of Technology, USA\\
\addr $^{5}$ Program in Health Sciences and Technology, Massachusetts Institute of Technology, USA
}

\maketitle


\begin{abstract}
We present a deep learning strategy that enables, for the first time, contrast-agnostic semantic segmentation of completely unpreprocessed  brain MRI scans, without requiring additional training or fine-tuning for new modalities. 
Classical Bayesian methods address this segmentation problem with unsupervised intensity models, but require significant computational resources. In contrast, learning-based methods can be fast at test time, but are sensitive to the data available at training. Our proposed learning method, \netname{}, leverages a set of training segmentations (no intensity images required) to generate synthetic scans of widely varying contrasts on the fly during training. These scans are produced using the generative model of the classical Bayesian segmentation framework, with randomly sampled parameters for appearance, deformation, noise, and bias field. Because each mini-batch has a different synthetic contrast, the final network is not biased towards any MRI contrast.
We comprehensively evaluate our approach on four datasets comprising over 1,000 subjects and four MR contrasts. The results show that our approach successfully segments every contrast in the data, performing slightly better than classical Bayesian segmentation, and three orders of magnitude faster. Moreover, even within the same type of MRI contrast, our strategy generalizes significantly better across datasets, compared to training using real images. Finally, we find that synthesizing a broad range of contrasts, even if unrealistic, increases the generalization of the neural network. Our code and model are open source at~\url{https://github.com/BBillot/\netname}.
\end{abstract}



\section{Introduction}

Segmentation of brain MR scans is an important task in neuroimaging, as it is a primary step in a wide array of subsequent analyses such as volumetry, morphology, and connectivity studies. Despite the success of modern supervised segmentation methods, especially convolutional neural networks (CNN), their adoption in neuroimaging has been hindered by the high variety in MRI contrasts. These approaches often require a large set of manually segmented preprocessed images \textit{for each} desired contrast. However, since manual segmentation is costly, such supervision is often not available.  A straightforward solution, implemented by widespread neuroimaging packages like FreeSurfer \cite{fischl_freesurfer_2012} or FSL \cite{jenkinson_fsl_2012}, is to require a 3D, T1-weighted scan for every subject, which is aggressively preprocessed, then used for segmentation purposes. However, such a requirement precludes analysis of datasets for which 3D T1 scans are not available.

Robustness to MRI contrast variations has classically been achieved with Bayesian methods. These approaches rely on a generative model of brain MRI scans, which combines an anatomical prior (a statistical atlas) and a likelihood distribution. The likelihood typically models the image intensities of different brain regions as a Gaussian mixture model (GMM), as well as artifacts such as bias field. Test scans are segmented by ``inverting'' this  generative model using Bayesian inference. 
If the GMM parameters are independently derived from each test scan in an unsupervised fashion \cite{van_leemput_automated_1999,zhang_segmentation_2001,ashburner_unified_2005}, this approach is fully adaptive to any MRI contrast. In some cases, \emph{a priori} information is included in the parameters, which constrains the method to a specific contrast \cite{wells_adaptive_1996,fischl_whole_2002,patenaude_bayesian_2011} -- yet even these methods are generally robust to small contrast variations.  Such robustness is an important reason why Bayesian techniques remain at the core of all major neuroimaging packages, such as FreeSurfer, FSL, or SPM~\cite{ashburner_spm_2012}. However, these strategies require significant computational resources (tens of minutes per scan) compared to recent deep learning methods, limiting large-scale deployment or time-sensitive applications.

Another popular family of neuroimaging segmentation methods is multi-atlas segmentation (MAS)~\cite{rohlfing_evaluation_2004,iglesias_multi-atlas_2015}. In MAS, several labeled scans (``atlases'') are registered to the test scan, and their deformed labels are merged into a final segmentation with a label-fusion algorithm~\cite{sabuncu_generative_2010}. MAS was originally designed for intra-modality problems, but can be extended to cross-modality applications by using multi-modality registration metrics like mutual information~\cite{wells_adaptive_1996,maes_multimodality_1997}. However, their performance in this scenario is poor, due to the limited accuracy of nonlinear registration algorithms across modalities~\cite{iglesias_is_2013}. Another main drawback of MAS has traditionally been the high computational cost of the multiple nonlinear registrations. While this is quickly changing with the advent of fast, deep learning registration techniques~\cite{balakrishnan_voxelmorph_2019,de_vos_end--end_2017}, accurate deformable registration for arbitrary modalities has not been widely demonstrated with these methods.

The modern segmentation literature is dominated by CNNs~\cite{milletari_v-net_2016,kamnitsas_efficient_2017}, particularly the U-Net architecture~\cite{ronneberger_u-net_2015}. Although CNNs produce fast and accurate segmentations when trained for modality-specific applications, they typically do not generalize well to image contrasts which are different from the training data~\cite{akkus_deep_2017,jog_pulse_2018,karani_lifelong_2018}. A possible solution is to train a network with multi-modal data, possibly with modality dropout during training~\cite{havaei_hemis_2016}, although this assumes access to manually labeled data on a wide range of acquisitions, which is problematic. One can also augment the training dataset with synthetic contrast variations that are not initially available from uni- or multi-modal scans~\cite{chartsias_multimodal_2018,huo_synseg-net_2019,kamnitsas_unsupervised_2017,jog_pulse_2018}. Recent papers have also shown that spatial and intensity data augmentation can improve network robustness~\cite{chaitanya_semi-supervised_2019,zhao_data_2019}. Although these approaches make segmentation CNNs adaptive to brain scans of observed contrasts, they remain limited to the modalities (real or simulated) present in the training data, and thus have reduced accuracy when tested on previously unseen MR contrasts.

To address modality-agnostic learning-based segmentation, a CNN was recently used to quickly solve the inference problem within the Bayesian segmentation framework~\cite{dalca_unsupervised_2019}. However, this method cannot be directly used to segment test scans of arbitrary contrasts, as it requires training on a set of unlabeled scans for each target modality.

In this paper, we present \netname{}, a novel learning strategy that enables automatic segmentation of \textit{unpreprocessed} brain scans of \emph{any} MRI contrast without any need for paired training data, re-training, or fine tuning. We train a CNN using a dataset of only segmentation maps: synthetic images are produced by sampling a generative model of Bayesian segmentation, conditioned on a segmentation map. By sampling model parameters randomly at every mini-batch, we expose the CNN to synthetic (and often unrealistic) contrasts during training, and force it to learn features that are inherently contrast agnostic. Our experiments demonstrate \netname{} on four different MRI contrasts. We also show that even within the same MRI contrast, \netname{} generalizes across datasets better than a CNN trained on real images of this contrast.


\section{Methods}

We first introduce the generative model for Bayesian MRI segmentation, and then describe our method, which builds on this framework to achieve modality-agnostic segmentation.

\subsection{Classical generative model for Bayesian segmentation of brain MRI}
\label{bayesian}

The Bayesian segmentation framework relies on a probabilistic generative model for brain  scans. Let~$L$ be a 3D label (segmentation) map consisting of~$J$ voxels, such that each voxel value $L_j$ is one of~$K$ possible labels: $L_j \in \{1,\ldots,K\}$. The generative model starts with a prior anatomical distribution $p(L)$, typically represented as a (precomputed) statistical atlas~$A$, which associates each voxel location with a $K$-length vector of label probabilities. Additionally, the atlas $A$ is endowed with a spatial deformation model: the label probabilities are warped with a field $\phi$, parameterized by $\theta_{\phi}$, which follows a distribution~$p(\theta_{\phi})$ chosen to encourage smooth deformations. The probability of observing $L$ is then:
\begin{align}
    p(L| A, \theta_\phi) = \prod_{j=1}^{J} [A \circ \phi(\theta_{\phi})]_{j,L_j},
\end{align}
where $[A \circ \phi(\theta_{\phi})]_{j,L_j}$ is the probability of label $L_j$ given by the warped atlas at location $j$. 

Given a label map $L$, the image likelihood~$p(I|L)$ for its corresponding image $I$ is commonly modeled as a GMM (conditioned on $L$), modulated by smooth, multiplicative bias field noise (additive in the more convenient logarithmic domain). Specifically, each label $k \in \lbrace1,...,K\rbrace$ is associated with a Gaussian distribution for intensities of mean $\mu_{k}$, and standard deviation $\sigma_{k}$. We group these Gaussian parameters into $\theta_G = \{\mu_1,\sigma_1,\ldots,\mu_K,\sigma_K\}$. The bias field is often modeled as a linear combination of smooth basis functions, where linear coefficients are grouped in $\theta_{B}$ \cite{larsen_n3_2014}. The image likelihood is given by:
\begin{align}
    p(I | L, \theta_B, \theta_G) = \prod_j \mathcal{N}(I_j - B_j(\theta_B) ; \mu_{L_j} , \sigma_{L_j}^2),
    \label{eq:likelihood}
\end{align}
where~$\mathcal{N}(\cdot; \mu, \sigma^2)$ is the Gaussian distribution, $I_j$ is the image intensity at voxel $j$, and $B_j(\theta_B)$ is the bias field at voxel $j$. We assume that $I_j$ has been log-transformed, such that the bias field is additive, rather than multiplicative.

Bayesian segmentation uses Bayes's rule to ``invert" this generative model to estimate~$p(L|I)$, posing segmentation as an optimization problem. Such inversion often relies on computing point estimates for the model parameters. Fitting the Gaussian parameters $\theta_G$ to the intensity distribution of the test scan is what makes these methods contrast agnostic.

\begin{figure}[t]
\centering
\floatconts
  {fig:schematic}
  {
  \caption{\netname{} overview. The proposed data generation process selects one of the available label maps~$S_m$ and employs a sampling strategy to synthesize an image-segmentation pair~$\{I,L\}$, based on a well-established generative model of brain MRI. Specific generation steps are illustrated in Figure~\ref{fig:augm_example} and detailed in Algorithm~\ref{alg:net}. The pairs~$\{I,L\}$ are used to train a CNN in a supervised fashion.}}
  {\includegraphics[width=1\textwidth]{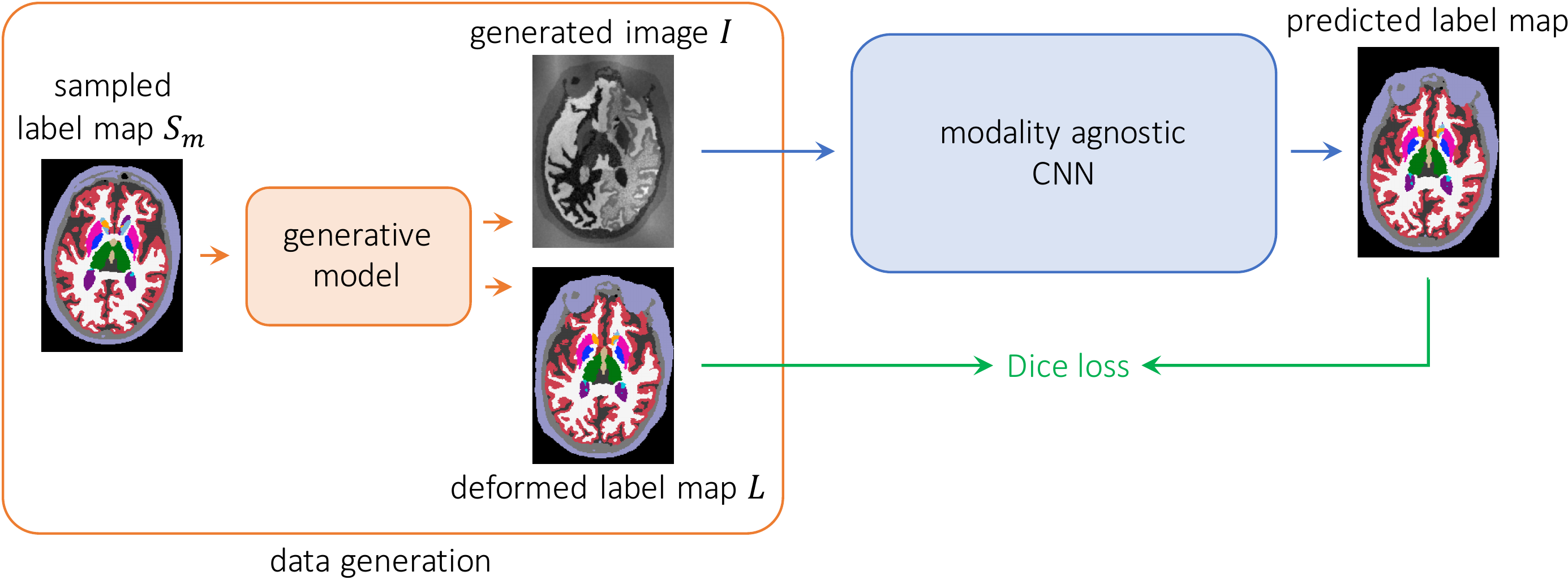} 
  }
\end{figure}
\begin{algorithm2e}[t]
\caption{Proposed Learning Strategy for \netname{}}
\label{alg:net}
\DontPrintSemicolon
\KwIn{$\{S_m\}_{m=1,\ldots,M}$   \tcp*{M segmentations}}
\While{not converged}{
$i \sim \mathcal{U}_d(1,M)$  \tcp*{select input map}
$\theta_{aff}  \sim \mathcal{U}(a_{rot},b_{rot})  \times   \mathcal{U}(a_{sc},b_{sc})  \times  \mathcal{U}(a_{sh},b_{sh}) \times \mathcal{U}(a_{tr},b_{tr}) $  \tcp*{affine parameters}
$\theta_v \sim \mathcal{N}_{10\times10\times10\times3}(0,\sigma_{svf}^2)$ \tcp*{sample SVF parameters}
$\phi_v(\theta_v) \leftarrow ScaleAndSquare[Upscale(\theta_v)]$ \tcp*{upscaling and integration}
$\phi \leftarrow \phi_{aff}(\theta_{aff}) \circ \phi_v(\theta_v)$  \tcp*{form deformation}
$L \leftarrow S_i \circ \phi$  \tcp*{deform selected label map}
$(\mu_k,\sigma_k) \sim  \mathcal{U}(a_{\mu},b_{\mu}) \times \mathcal{U}(a_{\sigma},b_{\sigma}), k=1,\ldots,K$  \tcp*{sample Gaussian parameters}
$G_j \sim \mathcal{N}(\mu_{L_{ij}},\sigma_{L_{ij}})$  \tcp*{sample GMM image}
$G^{blur} \leftarrow G * R(\sigma_{blur})$ \tcp*{Spatial blurring}
$\theta_B \sim \mathcal{N}_{4\times4\times4}(0,\sigma_{b}^2)$ \tcp*{sample bias field parameters}
$B  \leftarrow \exp[Upscale(\theta_B)]$ \tcp*{upscaling and exponentiation}
$G^{bias} \leftarrow G^{blur} \odot B$ \tcp*{bias field corruption}
$\gamma \sim   \mathcal{U}(a_\gamma,b_\gamma)$ \tcp*{gamma augmentation parameter}
$I \leftarrow f(G^{bias},\gamma)$  \tcp*{gamma and normalization via~\eqref{eq:intensity_augm}}
update CNN weights with pair $\{I, L\}$  \tcp*{SGD iteration}
}
\end{algorithm2e}

\subsection{Proposed approach}
\label{sec:approach}

We propose to train a segmentation CNN using synthetic data created on the fly with a generative model similar to that of Bayesian segmentation. Since the voxel independence assumption would yield extremely heterogeneous noisy images, we rely on a set of $M$ original label maps~$S=\{S_m\}_{m=1}^M$ instead of random samples from a probabilistic atlas. We also slightly blur the sampled intensities. The proposed learning strategy, detailed below, is summarized in \figureref{fig:schematic} and Algorithm~\ref{alg:net}, and exemplified in \figureref{fig:augm_example}. \newline

\paragraph{Data sampling:} In training, mini-batches are created  by sampling image-segmentation pairs~$\{I,L\}$ as follows. First, we randomly select a label map $S_i$ from the training dataset (\figureref{fig:augm_example}a), by sampling $i \sim \mathcal{U}_d(1,M)$, where $\mathcal{U}_d$ is the discrete uniform distribution. 

\begin{figure}[t]
\centering
\floatconts
  {fig:augm_example}
  {
  \caption{Intermediate steps of image generation (axial slices of 3D volumes). (a)~Segmentation. (b)~Warp with random smooth deformation field. (c)~Image intensities sampled via a GMM with random parameters. (d)~Blur. (e)~Random  bias field. (f)~Synthesized images with the contours of the corresponding label maps.}}
  {\includegraphics[width=\textwidth]{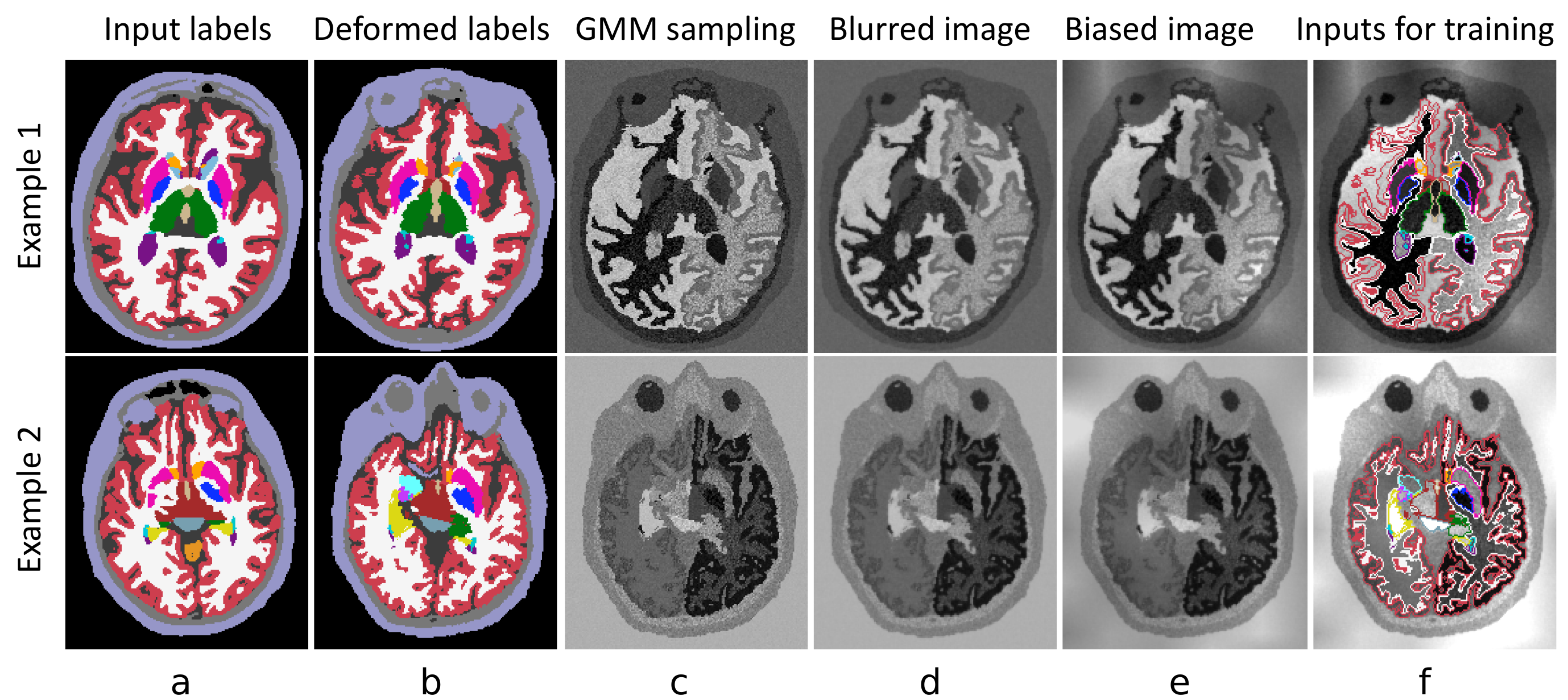}}
\end{figure}

Next, we generate a random deformation field~$\phi$ to obtain a new anatomical map~\mbox{$L = S_i \circ \phi$}. The deformation field $\phi$ is the composition of an affine and a deformable random transform, $\phi_{aff}$ and $\phi_{v}$, parameterized by $\theta_{aff}$ and $\theta_v$, respectively:~$\theta_\phi=(\theta_{aff},\theta_v)$. The affine component is the composition of three rotations ($\theta_{rot}$), three scalings ($\theta_{sc}$), three shears ($\theta_{sh}$), and three translations ($\theta_{tr}$). All these parameters are independently sampled from 
continuous uniform distributions with predefined ranges: $\mathcal{U}(a_{rot},b_{rot})$, 
$\mathcal{U}(a_{sc},b_{sc})$,
$\mathcal{U}(a_{sh},b_{sh})$, 
and $\mathcal{U}(a_{tr},b_{tr})$, respectively.
The deformable component is a diffeomorphic transform, obtained by integrating a smooth, random stationary velocity field (SVF) with a scaling and squaring approach~\cite{moler_nineteen_2003,arsigny_log-euclidean_2006}, implemented efficiently for a GPU~\cite{dalca_unsupervised_2019-1,krebs_learning_2019}. The SVF is generated by first sampling the parameters $\theta_v$. This is a random, low-resolution tensor (size $c_v \times c_v \times c_v \times3$), where each element is a sample from a zero-mean Gaussian distribution with standard deviation $\sigma_{svf}$. This tensor is subsequently upscaled to the desired image resolution with trilinear interpolation, to obtain a smooth SVF, which is  integrated to obtain $\phi_v$. The final deformed label map is obtained by resampling
\begin{align}
L = S_i \circ \phi =  S_i \circ [\phi_{aff}(\theta_{aff})\circ \phi_v(\theta_v)]
\end{align}
with nearest neighbor interpolation. This generative model yields a wide distribution of neuroanatomical shapes, while ensuring spatial smoothness (\figureref{fig:augm_example}b).

Given the segmentation $L$, we sample a synthetic image $I$ as follows. First, we sample an image $G$ conditioned on $L$, following the likelihood model introduced in section \ref{bayesian}, one voxel at the time using~$G_j \sim \mathcal{N}(\mu_{L_j}, \sigma_{L_j}^2)$. The Gaussian parameters~$\{\mu_k, \sigma_k\}$ are a set of $K$ independent means and standard deviations drawn from continuous uniform distributions $\mathcal{U}(a_\mu,b_\mu)$ and $\mathcal{U}(a_\sigma,b_\sigma)$, respectively. Sampling independently from a wide range of values yields images of extremely diverse contrasts (\figureref{fig:augm_example}c). To mimic  partial volume effects, we make the synthetic images more realistic by introducing a small degree of spatial correlation between neighboring voxels. This is achieved by blurring $G$ with a Gaussian kernel $R(\sigma_{blur})$ with standard deviation $\sigma_{blur}$ voxels, i.e., $G^{blur} = G * R(\sigma_{blur})$ (\figureref{fig:augm_example}d).

We corrupt the images with a bias field $B$, parameterized by $\theta_B$.  $B$ is generated in a similar way as the SVF: $\theta_B$ is a random, low resolution tensor (size $c_B\times c_B \times c_B$ in our experiments), whose elements are independent samples of a Gaussian distribution $\mathcal{N}(0,\sigma_b)$. This tensor is upscaled to the image size of $L$ with trilinear interpolation, and the voxel-wise exponential is taken to ensure non-negativity. The bias field corrupted image $G^{bias}$ is obtained by voxel-wise multiplication: $G^{bias} = G^{blur} \odot B$ (\figureref{fig:augm_example}e). 

Finally, the training image $I$ is generated by standard gamma augmentation and normalization of intensities. We first sample $\gamma$  from a uniform distribution $\mathcal{U}(a_\gamma,b_\gamma)$ and then:
\begin{equation}
\label{eq:intensity_augm}
I_j = \left( [G^{bias}_j - \min_{j}(G^{bias}_j)] \bigg/ [\max_{j}(G^{bias}_j)-\min_{j}(G^{bias}_j)] \right)^{\gamma}.
\end{equation}

\paragraph{Training:} Starting from a set of label maps, we use the generative process described above to form training pairs~$\{I, L\}$ (\figureref{fig:augm_example}f). These pairs -- each sampled with different parameters -- are used to train the CNN in a standard supervised fashion~(\figureref{fig:schematic}).

\begin{table}[tbp]
\setlength\tabcolsep{3pt} 
\floatconts
  {tab:hyperparameters}
  {\caption{Hyperparameters used in our experiments. Angles are in degrees;  spatial measures are in voxels. Intensity hyperparameters assume an input in the [0,255] interval.}  
  }
  {\small \begin{tabular}{|c|c|c|c|c|c|c|c|c|c|c|c|c|c|c|c|c|c|c|}
  \hline
 $a_{rot}$ & $b_{rot}$ & $a_{sc}$ & $b_{sc}$ & $a_{sh}$ & $b_{sh}$ & $a_{tr}$ & $b_{tr}$ & $\sigma_{svf}$ & $a_\mu$ & $b_\mu$ & $a_\sigma$ & $b_\sigma$ &$\sigma_{blur}$ & $\sigma_b$ & $a_\gamma$ & $b_\gamma$ & $c_v$ & $c_B$ \\
  \hline
  -10 & 10 & 0.9 & 1.1 & -0.01 & 0.01 & -20 & 20 & 3 & 25 & 225 & 5 & 25 & 0.3 & 0.5 & -0.3 & 0.3 & 10 & 4\\
  \hline
  \end{tabular}}
\end{table}

\subsection{Implementation details}
\label{implementation details}

\textbf{Architecture:} We use a U-Net style architecture~\cite{ronneberger_u-net_2015,cicek_3d_2016} with 5 levels of 2 layers each. The first layer contains 24 feature maps, and this number is doubled after each max-pooling, and halved after each upsampling. Convolutions are performed with kernels of size $3\times3\times3$, and use the Exponential Linear Unit as activation function~\cite{clevert_fast_2016}. We also make use of batch-normalization layers before each max-pooling and upsampling layer~\cite{ioffe_batch_2015}. The last layer uses a softmax activation function. The loss function is the average soft Dice~\cite{milletari_v-net_2016} coefficient between the ground truth segmentation and the probability map corresponding to the predicted output.

\paragraph{Parametric distributions and intensity constraints:}
The proposed generative model involves several hyperparameters (described above), which control the priors of model parameters. 
In order to achieve invariance to input contrast, we sample the hyperparameters of the GMM (describing priors for intensity means and variances)  from wide ranges in an independent fashion, generally leading to  unrealistic images (\figureref{fig:augm_example}). The deformation hyperparameters are chosen to yield a wide range of shapes -- well beyond plausible anatomy. We emphasize that the hyperparameter values, summarized in \tableref{tab:hyperparameters}, are \textit{not} chosen to mimic a particular imaging modality or subject cohort. 

\paragraph{Skull stripping:} 
The proposed method is designed to segment brain MRI without any preprocessing. However, in practice, some brain MRI datasets do not include extracerebral tissue, for example due to privacy issues. We build robustness to skull-stripped images into our method, by treating all extracerebral regions as background in 20\% of training samples. 

\paragraph{GPU implementation:}Our model, including the image sampling process, is~\mbox{implemented} on the GPU in Keras \cite{chollet_keras_2015} with a Tensorflow backend \cite{abadi_tensorflow_2016}.


\section{Experiments and results}

We provide experiments to evaluate segmentation of \textit{unprocessed} scans, eliminating the dependence on additional tools which can be CPU intensive and require manual tuning.

\subsection{Datasets}

We use four datasets with an array of modalities, and contrast variations within modalities. All datasets contain labels for 37 regions of interest (ROIs), with the same labeling protocol.

\paragraph{T1-39:} 39 whole head T1 scans with manual segmentations \cite{fischl_freesurfer_2012}. We split the dataset into subsets of 20 and 19 scans. We use the labels maps of the first 20 as the only inputs to train \netname{}, and evaluate on the held-out 19. We augmented the manual labels with approximate segmentations for skull, eye fluid, and other extra-cerebral tissue, computed semi-automatically with in-house tools, to enable synthesis of full head scans.

\paragraph{T1mix:}  1,000 T1 whole head MRI scans collected from seven public datasets: ABIDE \cite{di_martino_autism_2014}, ADHD200 \cite{the_adhd-200_consortium_adhd-200_2012}, GSP \cite{holmes_brain_2015}, HABS \cite{dagley_harvard_2017}, MCIC \cite{gollub_mcic_2013}, OASIS \cite{marcus_open_2007}, and PPMI \cite{marek_parkinson_2011}. Although these scans share the same modality, they exhibit variability in intensity distributions and head positioning due to differences in acquisition platforms and sequences. Since manual delineations are not available for these scans, we evaluate against automated segmentations obtained with FreeSurfer~\cite{fischl_freesurfer_2012,dalca_anatomical_2018}. T1mix enables evaluation on a large dataset of heterogeneous T1 contrasts.

\paragraph{FSM:} 18 subjects, each with 3 modalities: T1, T2, and a sequence typically used in deep brain stimulation (DBS). The DBS scan is an MP-RAGE with: TR $= \SI{3000}{\milli\second}$, TI $= \SI{406}{\milli\second}$, TE $=\SI{3.56}{\milli\second}$, $\alpha = 8^\circ$. With no manual delineations available, for evaluation we use automated segmentations produced by FreeSurfer on the T1 channel as ground truth for all modalities. This dataset enables evaluation on two new contrasts, T2 and DBS.

\paragraph{T1-PD-8:} T1 and proton density (PD) scans for 8 subjects, with manual delineations. These scans were approximately skull stripped prior to availability. Despite its smaller size, this dataset enables evaluation on another contrast (PD) that is very different than T1. \newline

\noindent Although FreeSurfer segmentations are not as accurate as manual delineations, they enable evaluation where manual labels are missing. FreeSurfer has been thoroughly evaluated on numerous independent datasets~\cite{fischl_whole_2002, tae_validation_2008}. It also yields high Dice scores against manual segmentations for T1-39 (0.88, albeit biased by mixing FreeSurfer training and testing data) and T1-PD-8 (0.85).

\subsection{Competing methods}

We compare our method \netname{} with three other approaches:

\paragraph{Fully supervised network:} We train a \textit{supervised} U-Net on the 20 training images from the T1-39 dataset (whole brain, unprocessed), aiming to assess difference in performance when testing on images of the same contrast (T1) acquired on the same and other platforms. 
We employ the same architecture and loss function as for \netname{}, and we use the same data augmentation, including spatial deformation, gamma augmentation (i.e., $y=x^\gamma$), and normalization of intensities. This supervised network can only segment T1 scans, so we refer to it as ``T1 baseline".

\paragraph{SAMSEG:}  Based on the traditional Bayesian segmentation framework, SAMSEG~\cite{puonti_fast_2016} uses unsupervised likelihood distributions, and is thus fully contrast-adaptive. Like our method, SAMSEG can segment both unprocessed or skull-stripped scans. SAMSEG does not rely on neural networks, and thus does not require training, but instead employs an independent optimization for each scan requiring tens of minutes.

\paragraph{\netname{}-rule:} We also analyze a variant of our proposed method, where the intensity parameters are representative of the test scans to be segmented. For each of the seven contrasts present in the training data (T2, PD, DSB, and four varieties of T1), we build a Gaussian hyperprior for the means and standard deviations of each label, using ground truth segmentations. At training, for every mini-batch we sample one of the seven contrasts, then we sample the means and standard deviations for each class conditioned on the contrast. This variant enables us to compare the generation of unrealistic contrasts during training, against enforcing prior information on the target modalities, if available. An example of these more realistic synthetic images (conditioned on T1 contrast) is shown in \figureref{fig:augm_realistic}.

\begin{figure}[t]
\floatconts
  {fig:augm_realistic}
  {\caption{Generation of a T1-like image for training \netname{}-rule.}}
  {\centering\includegraphics[width=0.90\textwidth]{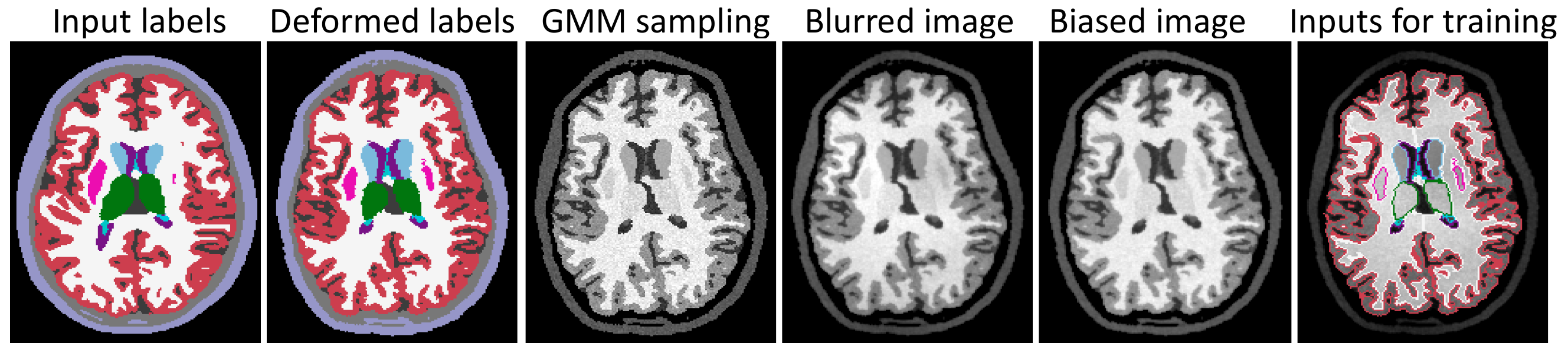}}
\end{figure}

\subsection{Experimental setup}

All CNN methods are trained on the training subset of T1-39, with our method variants only requiring the segmentation maps, whereas the supervised baseline also uses the T1 scans. We evaluate all approaches on the test subset of T1-39, as well as all of T1mix, T1-PD-8, and FSM. The T1 baseline is not tested on modalities other than T1, nor on T1-PD-8 because it cannot cope with skull stripped data. We assess performance with Dice scores, computed for a representative subset of 12 brain ROIs: cerebral white matter (WM) and cortex (CT), lateral ventricle (LV), cerebellar white matter (CW) and cortex (CC), thalamus (TH), caudate (CA); putamen (PU), pallidum (PA), brainstem (BS), hippocampus (HP), and amygdala (AM). We averaged results for contralateral structures.

\subsection{Results}

\begin{table}[tbp]
\setlength\tabcolsep{3pt} 
\floatconts
  {tab:summary}
  {\caption{Summary of results, capturing the performance of each method, its ability to segment arbitrary modalities, and run time (averaged over 10 runs). SAMSEG was run using 8 cores (Intel Xeon at 3.00GHz), whereas \netname{} was run on an Nvidia P6000 GPU. Image loading time was not considered.}}
  {\small \begin{tabular}{c || c |  c |  c}
  \hline
  Method & Overall performance & modality-agnostic & runtime (s)  \\
  \hline \hline
 Supervised & 0.89 $\pm$ 0.10 (same dataset) 0.59 $\pm$ 0.11 (other T1s) & No & 3.06 $\pm$ 0.02\\  
 SAMSEG & 0.83 $\pm$ 0.02  & Yes & 1382 $\pm$ 192\\  
 \netname{}-rule & 0.82 $\pm$ 0.02 & Yes & 3.22 $\pm$ 0.03\\  
 \netname{} & 0.85 $\pm$ 0.02 & Yes & 3.22 $\pm$ 0.03\\  
  \hline
  \end{tabular}}
\end{table}

\tableref{tab:summary} provides a summary of the methods and their runtime. \figureref{fig:dice} shows box plots for each ROI, method, and dataset, as well as averages across the ROIs. Table~\ref{tab:pvalues} shows corresponding median scores and p values using Wilcoxon test. Finally,  \figureref{fig:examples} shows sample segmentations for every method and dataset. The supervised T1 baseline excels when tested on the test scans of T1-39 (i.e., intra-dataset), achieving a mean Dice of 0.89, and outperforming all the other methods for every ROI. However, when tested on T1 images from T1mix and FSM, we observe substantial variations in Dice scores (e.g., across the different sub-datasets within T1mix), with a consistent decrease in performance (see for instance the segmentation of the T1 in FSM in \figureref{fig:examples}). This is likely due to the limited variability in the training dataset, despite the use of augmentation techniques, highlighting the challenge of variation in \textit{unprocessed} scans from different sources, even within the same modality.

\begin{figure}[t]
\floatconts
  {fig:dice}
  {\caption{Dice scores obtained by each method, shown both in aggregate (top left) and per individual ROIs for each dataset. Sub-datasets of T1mix are marked with a star.}}
  {\includegraphics[width=\textwidth]{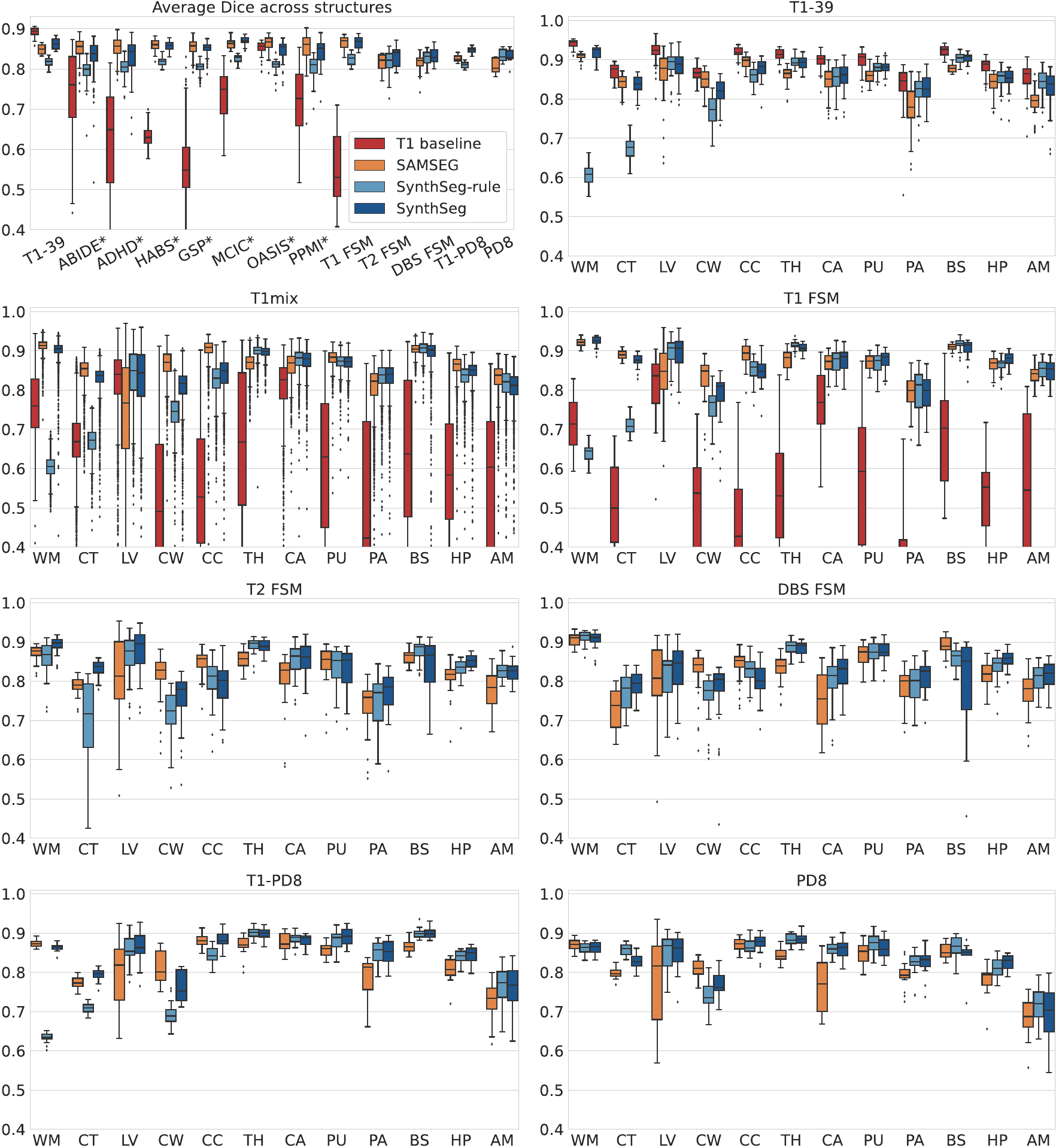}}
\end{figure}

\begin{table}[tbp]
\setlength\tabcolsep{3pt} 
\floatconts
  {tab:pvalues}
  {\caption{Median Dice scores and p values for two-sided non-parametric Wilcoxon signed-rank tests comparing \netname{} and the competing methods. Sub-datasets of T1mix are marked with a star.}}
  {\small \begin{tabular}{|c|c|c|c|c|c|c|c|}
  \hline
  & \netname{}  & \multicolumn{2}{c|}{T1-baseline}  &  \multicolumn{2}{c|}{SAMSEG} &   \multicolumn{2}{c|}{\netname{}-rule}  \\ \cline{2-8}
  Dataset  & Med. Dice  &  Med. Dice & p value  &   Med. Dice &  p value  &  Med. Dice  &  p value   \\ 
  \hline
  T1-39   & 0.861 & 0.894 & $p<10^{-3}$ &   {0.849} & $p<10^{-3}$  &  {0.819} & $p<10^{-3}$    \\ 
  T1mix   & 0.852 &  {0.601} & $p<10^{-94}$  & 0.858 & $p<10^{-30}$ & {0.806} & $p<10^{-85}$  \\ 
  ABIDE*  & 0.838 & {0.761} & $p<10^{-15}$  & 0.856 & $p<10^{-11}$   & {0.799} & $p<10^{-19}$  \\ 
  ADHD*   & 0.843  & {0.649} & $p<10^{-14}$  & 0.857 & $p<10^{-4}$   & {0.804} & $p<10^{-9}$  \\ 
  HABS*   & 0.858 & {0.630} & $p<10^{-4}$  & 0.859 & $0.7$  &  {0.819} & $p<10^{-4}$  \\ 
  GSP*    & 0.853  & {0.549} & $p<10^{-92}$  & 0.857 & $p<10^{-7}$  &  {0.806} & $p<10^{-90}$  \\ 
  MCIC*   & 0.869  & {0.750} & $p<10^{-5}$  & {0.863} & $ 6.4 \times 10^{-3}$  &  {0.829} & $p<10^{-5}$  \\ 
  OASIS*  & 0.855  & 0.857 & $p<10^{-4}$  & 0.867 & $p<10^{-10}$  &  {0.812} & $p<10^{-12}$  \\ 
  PPMI*   & 0.851  & {0.726} & $p<10^{-12}$  & 0.861 & $p<10^{-6}$  &  {0.811} & $p<10^{-12}$  \\
  T1 FSM  & 0.869  & {0.531} & $p<10^{-13}$  & 0.869 & $0.6$  &  {0.827} & $p<10^{-3}$  \\
  T2 FSM  & 0.841   & N/A   & N/A & {0.822} & $ 1.2 \times 10^{-3}$  &  {0.822} & $ 1.5 \times 10^{-2}$  \\  
  DBS FSM & 0.828  & N/A & N/A  & {0.821} & $ 3.8 \times 10^{-3}$  &  0.831 & $ 0.2$  \\                 
  T1-PD8  & 0.848  & N/A & N/A  & {0.823} & $ 1.7 \times 10^{-2}$  &  {0.810} & $ 1.2 \times 10^{-2}$  \\  
  PD-PD8  & 0.830  & N/A  & N/A  & {0.801} & $ 3.6 \times 10^{-2}$  &  0.830 & $ 0.4$  \\   
  \hline
  \end{tabular}}
\end{table}

\begin{figure}[t]
\floatconts
  {fig:examples}
  {\caption{Example segmentations for each method and dataset. We selected the median subject in terms of Dice scores across ROIs and methods.}}
  {\includegraphics[width=\textwidth]{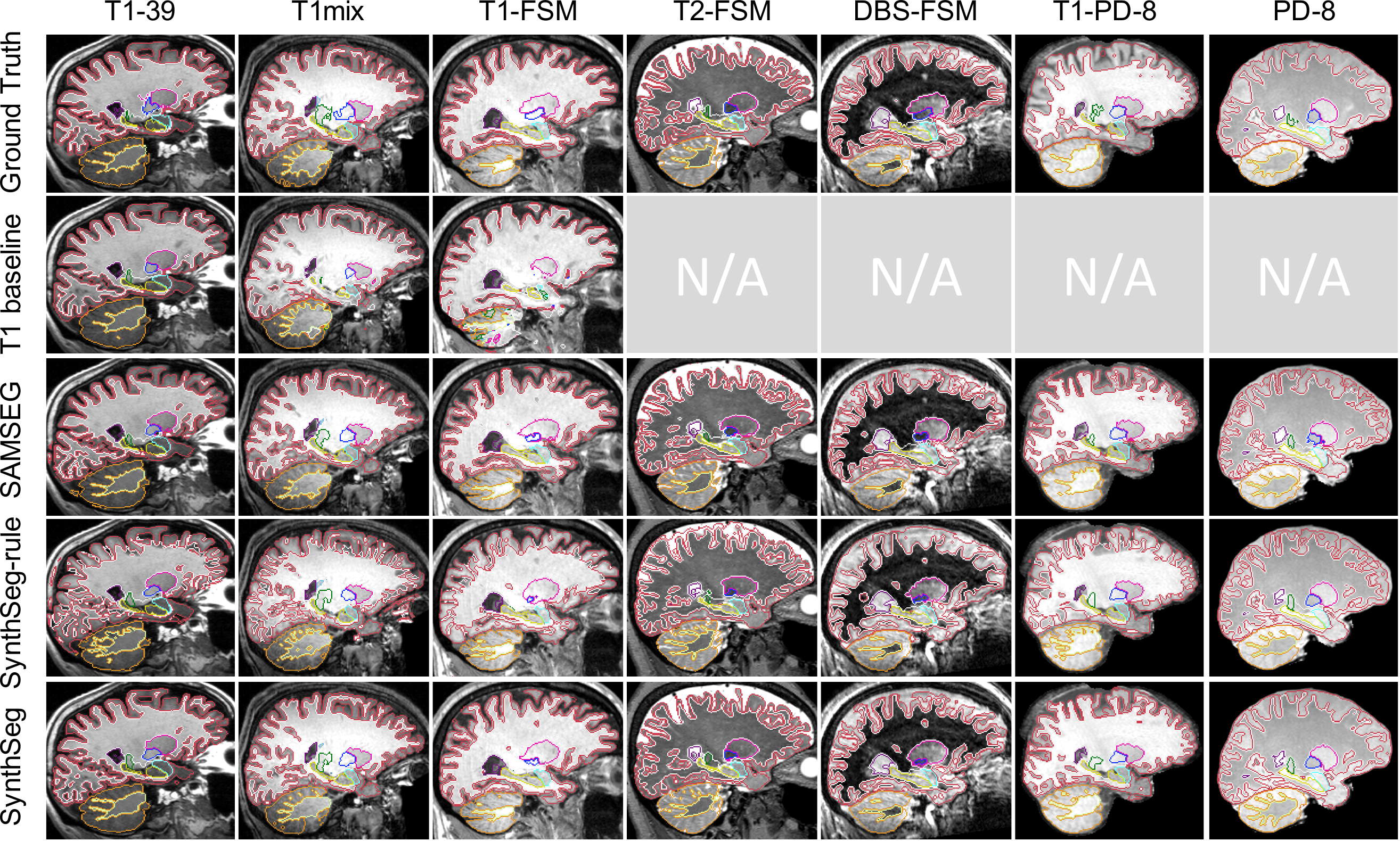}}
\end{figure}

SAMSEG yields very uniform results across datasets of T1 contrasts, producing mean Dice scores within 3 points of each other. Being agnostic to contrast, it outperforms the T1 baseline outside its training domain. It also performs well for the non-T1 contrasts. Although the mean Dice scores are slightly lower than for the T1 datasets (which normally display better contrast between gray and white matter), they remain robust for every contrast and dataset with minimum mean Dice of 0.81.

\netname{} also produces high Dice across all contrasts, slightly higher than SAMSEG (0.02 mean Dice improvement), while requiring a fraction of its runtime (\tableref{tab:summary}). The difference between SAMSEG and \netname{} is smaller for T1mix and T1 FSM because SAMSEG is positively biased by the use of FreeSurfer segmentations as ground truth, since the methods work similarly. The improvement of \netname{} compared to SAMSEG is consistent across structures, except the cerebellum. Compared to the T1 baseline, the mean Dice is 0.03 lower on the supervised training domain (T1-39), but generalizes significantly better to other T1 datasets, and can segment other MRI contrasts with little decrease in performance (minimum mean Dice is 0.83).

Importantly, \netname{}-rule is outperformed by \netname{}, and its Dice scores are also slightly lower than those produced by SAMSEG. This illustrates that adapting the parameters to a certain contrast is counterproductive, at least within our simple generative model: we observe consistent drops in performance across ROIs and datasets, despite injecting contrast-specific knowledge for each modality. This result is consistent  with recent results in image augmentation~\cite{chaitanya_semi-supervised_2019}, and supports the theory that forcing the network to learn to segment a broader range of images than it will typically observe at test time improves generalization.


\section{Discussion and conclusion}

We presented a learning strategy for modality-agnostic brain MRI segmentation, which builds on classical generative models for Bayesian segmentation. Sampling a wide range of model parameters enables the network to learn to segment a wide variety of contrasts and shapes during training. At test time, the network can therefore segment neuroanatomy given an unprocessed scan of any contrast in seconds. While the network is trained in a supervised fashion, the only data required are a few label maps. Importantly, we do not require any real scans during training, since images are synthesized from the labels, and are thus always perfectly aligned -- in contrast to techniques relying on manual delineations. Note that the presented method requires the training label maps to contain labels for all brain structures to be synthesized.

While a supervised network excels on test data from the same domain it was trained on, its performance quickly decays when faced with more variability, even within the same type of MRI contrast. We emphasize that this effect is particularly pronounced as we tackle the challenging task of segmentation starting with \textit{unprocessed} scans. This is one reason why deep learning segmentation techniques have not yet been adopted by widespread neuroimaging packages like FreeSurfer or FSL, where fewer assumptions on the specific MRI contrast of the user's data need to be made. In contrast, \netname{} maintains accuracy across T1 variants as well as other MRI modalities.

In absolute terms, \netname{}'s Dice scores are consistently high: higher than SAMSEG, and not far from \textit{supervised} contrast-specific networks, like the T1 baseline or scores reported in recent literature~\cite{roy_quicknat_2019}. Compared with our recent article that uses a CNN to estimate the GMM and registration parameters of the Bayesian segmentation framework~\cite{dalca_unsupervised_2019}, the method proposed here achieves higher average Dice on T1 (0.86 vs 0.82) and PD datasets (0.83 vs 0.80). However, we highlight that direct comparison is not available due to differences in datasets: in this work, we could only use 19 subjects from T1-39 for evaluation. More importantly, our previous method requires significant preprocessing and modality-specific unsupervised re-training. This highlights the ability of our new method to segment any contrasts without retraining or preprocessing; the latter eliminates the dependence on additional tools which can be computationally expensive and require manual tuning.

We believe that the proposed learning strategy is applicable to many generative models from which sampling can yields sensible data, even beyond neuroimaging. By greatly increasing the robustness of fast segmentation CNNs to a wide variety of MRI contrast, without any need for retraining, \netname{} promises to enable adoption of deep learning segmentation techniques by the neuroimaging community.


\midlacknowledgments{This research was supported by the European Research Council (ERC Starting Grant 677697, project BUNGEE-TOOLS), and by the EPSRC-funded UCL Centre for Doctoral Training in Medical Imaging (EP/L016478/1) and the Department of Health’s NIHR-funded Biomedical Research Centre at University College London Hospitals. 
Further support  was provided in part by the BRAIN Initiative Cell Census Network grant U01-MH117023, the National Institute for Biomedical Imaging and Bioengineering (P41-EB015896, 1R01-EB023281, R01-EB006758, R21-EB018907, R01-EB019956), the National Institute on Aging (1R56-AG064027, 5R01-AG008122, R01-AG016495, 1R01-AG064027), the National Institute of Mental Health  the National Institute of Diabetes and Digestive and Kidney Diseases (1-R21-DK-108277-01), the National Institute for Neurological Disorders and Stroke (R01-NS112161, R01-NS0525851, R21-NS072652, R01-NS070963, R01-NS083534, 5U01-NS086625,5U24-NS10059103, R01-NS105820), and was made possible by the resources provided by Shared Instrumentation Grants 1S10-RR023401, 1S10-RR01-9307, and 1S10-RR023043. Additional support was provided by the NIH Blueprint for Neuroscience Research (5U01-MH093765), part of the multi-institutional Human Connectome Project. In addition, BF has a financial interest in CorticoMetrics, a company whose medical pursuits focus on brain imaging and measurement technologies. BF's interests were reviewed and are managed by Massachusetts General Hospital and Partners HealthCare in accordance with their conflict of interest policies.}



\begin{thebibliography}{50}
\providecommand{\natexlab}[1]{#1}
\providecommand{\url}[1]{\texttt{#1}}
\expandafter\ifx\csname urlstyle\endcsname\relax
  \providecommand{\doi}[1]{doi: #1}\else
  \providecommand{\doi}{doi: \begingroup \urlstyle{rm}\Url}\fi

\bibitem[Abadi et~al.(2016)Abadi, Barham, Chen, Chen, Davis, Dean, Devin,
  Ghemawat, Irving, Isard, and {others}]{abadi_tensorflow_2016}
Martín Abadi, Paul Barham, Jianmin Chen, Zhifeng Chen, Andy Davis, Jeffrey
  Dean, Matthieu Devin, Sanjay Ghemawat, Geoffrey Irving, Michael Isard, and
  {others}.
\newblock Tensorflow: {A} system for large-scale machine learning.
\newblock In \emph{12th \$\{\${USENIX}\$\}\$ {Symposium} on {Operating}
  {Systems} {Design} and {Implementation} (\$\{\${OSDI}\$\}\$ 16)}, pages
  265--283, 2016.

\bibitem[Akkus et~al.(2017)Akkus, Galimzianova, Hoogi, Rubin, and
  Erickson]{akkus_deep_2017}
Zeynettin Akkus, Alfiia Galimzianova, Assaf Hoogi, Daniel Rubin, and Bradley
  Erickson.
\newblock Deep {Learning} for {Brain} {MRI} {Segmentation}: {State} of the
  {Art} and {Future} {Directions}.
\newblock \emph{Journal of Digital Imaging}, 30\penalty0 (4):\penalty0
  449--459, August 2017.
\newblock ISSN 1618-727X.
\newblock \doi{10.1007/s10278-017-9983-4}.

\bibitem[Arsigny et~al.(2006)Arsigny, Commowick, Pennec, and
  Ayache]{arsigny_log-euclidean_2006}
Vincent Arsigny, Olivier Commowick, Xavier Pennec, and Nicholas Ayache.
\newblock A log-{Euclidean} framework for statistics on diffeomorphisms.
\newblock \emph{Conference on Medical Image Computing and Computer-Assisted
  Intervention}, 9\penalty0 (Pt 1):\penalty0 924--931, 2006.
\newblock \doi{10.1007/11866565_113}.

\bibitem[Ashburner(2012)]{ashburner_spm_2012}
John Ashburner.
\newblock {SPM}: a history.
\newblock \emph{Neuroimage}, 62\penalty0 (2):\penalty0 791--800, 2012.
\newblock Publisher: Elsevier.

\bibitem[Ashburner and Friston(2005)]{ashburner_unified_2005}
John Ashburner and Karl Friston.
\newblock Unified segmentation.
\newblock \emph{NeuroImage}, 26\penalty0 (3):\penalty0 839--851, July 2005.
\newblock ISSN 1053-8119.
\newblock \doi{10.1016/j.neuroimage.2005.02.018}.

\bibitem[Balakrishnan et~al.(2019)Balakrishnan, Zhao, Sabuncu, Guttag, and
  Dalca]{balakrishnan_voxelmorph_2019}
Guha Balakrishnan, Amy Zhao, Mert Sabuncu, John Guttag, and Adrian Dalca.
\newblock {VoxelMorph}: {A} {Learning} {Framework} for {Deformable} {Medical}
  {Image} {Registration}.
\newblock \emph{IEEE Transactions on Medical Imaging}, 38\penalty0
  (8):\penalty0 1788--1800, August 2019.
\newblock ISSN 0278-0062, 1558-254X.
\newblock \doi{10.1109/TMI.2019.2897538}.

\bibitem[Chaitanya et~al.(2019)Chaitanya, Karani, Baumgartner, Becker, Donati,
  and Konukoglu]{chaitanya_semi-supervised_2019}
Krishna Chaitanya, Neerav Karani, Christian~F. Baumgartner, Anton Becker,
  Olivio Donati, and Ender Konukoglu.
\newblock Semi-supervised and {Task}-{Driven} {Data} {Augmentation}.
\newblock In \emph{Information {Processing} in {Medical} {Imaging}}, Lecture
  {Notes} in {Computer} {Science}, pages 29--41, Cham, 2019. Springer
  International Publishing.
\newblock ISBN 978-3-030-20351-1.
\newblock \doi{10.1007/978-3-030-20351-1_3}.

\bibitem[Chartsias et~al.(2018)Chartsias, Joyce, Giuffrida, and
  Tsaftaris]{chartsias_multimodal_2018}
Agisilaos Chartsias, Thomas Joyce, Mario Giuffrida, and Sotirios Tsaftaris.
\newblock Multimodal {MR} {Synthesis} via {Modality}-{Invariant} {Latent}
  {Representation}.
\newblock \emph{IEEE Transactions on Medical Imaging}, 37\penalty0
  (3):\penalty0 803--814, March 2018.
\newblock ISSN 1558-254X.
\newblock \doi{10.1109/TMI.2017.2764326}.

\bibitem[Chollet(2015)]{chollet_keras_2015}
François Chollet.
\newblock \emph{Keras (https://github.com/fchollet/keras)}.
\newblock 2015.

\bibitem[Clevert et~al.(2016)Clevert, Unterthiner, and
  Hochreiter]{clevert_fast_2016}
Djork-Arné Clevert, Thomas Unterthiner, and Sepp Hochreiter.
\newblock Fast and {Accurate} {Deep} {Network} {Learning} by {Exponential}
  {Linear} {Units} ({ELUs}).
\newblock \emph{arXiv:1511.07289 [cs]}, February 2016.

\bibitem[Consortium(2012)]{the_adhd-200_consortium_adhd-200_2012}
The ADHD-200 Consortium.
\newblock The {ADHD}-200 {Consortium}: {A} {Model} to {Advance} the
  {Translational} {Potential} of {Neuroimaging} in {Clinical} {Neuroscience}.
\newblock \emph{Frontiers in Systems Neuroscience}, 6, September 2012.
\newblock ISSN 1662-5137.
\newblock \doi{10.3389/fnsys.2012.00062}.

\bibitem[Dagley et~al.(2017)Dagley, LaPoint, Huijbers, Hedden, McLaren,
  Chatwal, Papp, Amariglio, Blacker, Rentz, Johnson, Sperling, and
  Schultz]{dagley_harvard_2017}
Alexander Dagley, Molly LaPoint, Willem Huijbers, Trey Hedden, Donald McLaren,
  Jasmeer Chatwal, Kathryn Papp, Rebecca Amariglio, Deborah Blacker, Dorene
  Rentz, Keith Johnson, Reisa Sperling, and Aaron Schultz.
\newblock Harvard {Aging} {Brain} {Study}: dataset and accessibility.
\newblock \emph{NeuroImage}, 144\penalty0 (Pt B):\penalty0 255--258, January
  2017.
\newblock ISSN 1053-8119.
\newblock \doi{10.1016/j.neuroimage.2015.03.069}.

\bibitem[Dalca et~al.(2019{\natexlab{a}})Dalca, Yu, Golland, Fischl, Sabuncu,
  and Iglesias]{dalca_unsupervised_2019}
Adrian Dalca, Evan Yu, Polina Golland, Bruce Fischl, Mert Sabuncu, and
  Juan~Eugenio Iglesias.
\newblock Unsupervised {Deep} {Learning} for {Bayesian} {Brain} {MRI}
  {Segmentation}.
\newblock \emph{arXiv:1904.11319 [cs, eess]}, July 2019{\natexlab{a}}.

\bibitem[Dalca et~al.(2018)Dalca, Guttag, and Sabuncu]{dalca_anatomical_2018}
Adrian~V. Dalca, John Guttag, and Mert~R. Sabuncu.
\newblock Anatomical {Priors} in {Convolutional} {Networks} for {Unsupervised}
  {Biomedical} {Segmentation}.
\newblock In \emph{Anatomical {Priors} in {Convolutional} {Networks} for
  {Unsupervised} {Biomedical} {Segmentation}}, pages 9290--9299, 2018.

\bibitem[Dalca et~al.(2019{\natexlab{b}})Dalca, Balakrishnan, Guttag, and
  Sabuncu]{dalca_unsupervised_2019-1}
Adrian~V. Dalca, Guha Balakrishnan, John Guttag, and Mert Sabuncu.
\newblock Unsupervised learning of probabilistic diffeomorphic registration for
  images and surfaces.
\newblock \emph{Medical Image Analysis}, 57:\penalty0 226--236, October
  2019{\natexlab{b}}.
\newblock ISSN 1361-8415.
\newblock \doi{10.1016/j.media.2019.07.006}.

\bibitem[de~Vos et~al.(2017)de~Vos, Berendsen, Viergever, Staring, and
  Išgum]{de_vos_end--end_2017}
Bob de~Vos, Floris Berendsen, Max Viergever, Marius Staring, and Ivana Išgum.
\newblock End-to-{End} {Unsupervised} {Deformable} {Image} {Registration} with
  a {Convolutional} {Neural} {Network}.
\newblock In \emph{Deep {Learning} in {Medical} {Image} {Analysis} and
  {Multimodal} {Learning} for {Clinical} {Decision} {Support}}, Lecture {Notes}
  in {Computer} {Science}, pages 204--212, Cham, 2017. Springer International
  Publishing.
\newblock ISBN 978-3-319-67558-9.
\newblock \doi{10.1007/978-3-319-67558-9_24}.

\bibitem[Di~Martino et~al.(2014)Di~Martino, Yan, Li, Denio, Castellanos,
  Alaerts, Anderson, Assaf, Bookheimer, Dapretto, Deen, Delmonte, Dinstein,
  Ertl-Wagner, Fair, Gallagher, Kennedy, Keown, Keysers, Lainhart, Lord, Luna,
  Menon, Minshew, Monk, Mueller, Müller, Nebel, Nigg, O’Hearn, Pelphrey,
  Peltier, Rudie, Sunaert, Thioux, Tyszka, Uddin, Verhoeven, Wenderoth,
  Wiggins, Mostofsky, and Milham]{di_martino_autism_2014}
Adriana Di~Martino, Chao-Gan Yan, Qingyang Li, Erin Denio, Francisco
  Castellanos, Kaat Alaerts, Jeffrey Anderson, Michal Assaf, Susan Bookheimer,
  Mirella Dapretto, Ben Deen, Sonja Delmonte, Ilan Dinstein, Birgit
  Ertl-Wagner, Damien Fair, Louise Gallagher, Daniel Kennedy, Christopher
  Keown, Christian Keysers, Janet Lainhart, Catherine Lord, Beatriz Luna, Vinod
  Menon, Nancy Minshew, Christopher Monk, Sophia Mueller, Ralph-Axel Müller,
  Mary~Beth Nebel, Joel Nigg, Kirsten O’Hearn, Kevin Pelphrey, Scott Peltier,
  Jeffrey Rudie, Stefan Sunaert, Marc Thioux, Michael Tyszka, Lucina Uddin,
  Judith Verhoeven, Nicole Wenderoth, Jillian Wiggins, Stewart Mostofsky, and
  Michael Milham.
\newblock The {Autism} {Brain} {Imaging} {Data} {Exchange}: {Towards}
  {Large}-{Scale} {Evaluation} of the {Intrinsic} {Brain} {Architecture} in
  {Autism}.
\newblock \emph{Molecular psychiatry}, 19\penalty0 (6):\penalty0 659--667, June
  2014.
\newblock ISSN 1359-4184.
\newblock \doi{10.1038/mp.2013.78}.

\bibitem[Fischl(2012)]{fischl_freesurfer_2012}
Bruce Fischl.
\newblock {FreeSurfer}.
\newblock \emph{NeuroImage}, 62\penalty0 (2):\penalty0 774--781, August 2012.
\newblock ISSN 1053-8119.
\newblock \doi{10.1016/j.neuroimage.2012.01.021}.

\bibitem[Fischl et~al.(2002)Fischl, Salat, Busa, Albert, Dieterich, Haselgrove,
  van~der Kouwe, Killiany, Kennedy, Klaveness, Montillo, Makris, Rosen, and
  Dale]{fischl_whole_2002}
Bruce Fischl, David Salat, Evelina Busa, Marilyn Albert, Megan Dieterich,
  Christian Haselgrove, Andre van~der Kouwe, Ron Killiany, David Kennedy, Shuna
  Klaveness, Albert Montillo, Nikos Makris, Bruce Rosen, and Anders Dale.
\newblock Whole brain segmentation: automated labeling of neuroanatomical
  structures in the human brain.
\newblock \emph{Neuron}, 33\penalty0 (3):\penalty0 341--355, January 2002.
\newblock ISSN 0896-6273.
\newblock \doi{10.1016/s0896-6273(02)00569-x}.

\bibitem[Gollub et~al.(2013)Gollub, Shoemaker, King, White, Ehrlich, Sponheim,
  Clark, Turner, Mueller, Magnotta, O’Leary, Ho, Brauns, Manoach, Seidman,
  Bustillo, Lauriello, Bockholt, Lim, Rosen, Schulz, Calhoun, and
  Andreasen]{gollub_mcic_2013}
Randy Gollub, Jody Shoemaker, Margaret King, Tonya White, Stefan Ehrlich, Scott
  Sponheim, Vincent Clark, Jessica Turner, Bryon Mueller, Vince Magnotta,
  Daniel O’Leary, Beng Ho, Stefan Brauns, Dara Manoach, Larry Seidman, Juan
  Bustillo, John Lauriello, Jeremy Bockholt, Kelvin Lim, Bruce Rosen, Charles
  Schulz, Vince Calhoun, and Nancy Andreasen.
\newblock The {MCIC} collection: a shared repository of multi-modal, multi-site
  brain image data from a clinical investigation of schizophrenia.
\newblock \emph{Neuroinformatics}, 11\penalty0 (3):\penalty0 367--388, July
  2013.
\newblock ISSN 1539-2791.
\newblock \doi{10.1007/s12021-013-9184-3}.

\bibitem[Havaei et~al.(2016)Havaei, Guizard, Chapados, and
  Bengio]{havaei_hemis_2016}
Mohammad Havaei, Nicolas Guizard, Nicolas Chapados, and Yoshua Bengio.
\newblock {HeMIS}: {Hetero}-{Modal} {Image} {Segmentation}.
\newblock In \emph{Medical {Image} {Computing} and {Computer}-{Assisted}
  {Intervention} – {MICCAI} 2016}, Lecture {Notes} in {Computer} {Science},
  pages 469--477, Cham, 2016. Springer International Publishing.
\newblock ISBN 978-3-319-46723-8.
\newblock \doi{10.1007/978-3-319-46723-8_54}.

\bibitem[Holmes et~al.(2015)Holmes, Hollinshead, O’Keefe, Petrov, Fariello,
  Wald, Fischl, Rosen, Mair, Roffman, Smoller, and Buckner]{holmes_brain_2015}
Avram Holmes, Marisa Hollinshead, Timothy O’Keefe, Victor Petrov, Gabriele
  Fariello, Lawrence Wald, Bruce Fischl, Bruce Rosen, Ross Mair, Joshua
  Roffman, Jordan Smoller, and Randy Buckner.
\newblock Brain {Genomics} {Superstruct} {Project} initial data release with
  structural, functional, and behavioral measures.
\newblock \emph{Scientific Data}, 2\penalty0 (1):\penalty0 1--16, July 2015.
\newblock ISSN 2052-4463.
\newblock \doi{10.1038/sdata.2015.31}.

\bibitem[Huo et~al.(2019)Huo, Xu, Moon, Bao, Assad, Moyo, Savona, Abramson, and
  Landman]{huo_synseg-net_2019}
Yuankai Huo, Zhoubing Xu, Hyeonsoo Moon, Shunxing Bao, Albert Assad, Tamara
  Moyo, Michael Savona, Richard Abramson, and Bennett Landman.
\newblock {SynSeg}-{Net}: {Synthetic} {Segmentation} {Without} {Target}
  {Modality} {Ground} {Truth}.
\newblock \emph{IEEE Transactions on Medical Imaging}, 38\penalty0
  (4):\penalty0 1016--1025, April 2019.
\newblock ISSN 1558-254X.

\bibitem[Iglesias and Sabuncu(2015)]{iglesias_multi-atlas_2015}
Juan~Eugenio Iglesias and Mert~R Sabuncu.
\newblock Multi-atlas segmentation of biomedical images: a survey.
\newblock \emph{Medical image analysis}, 24\penalty0 (1):\penalty0 205--219,
  2015.
\newblock Publisher: Elsevier.

\bibitem[Iglesias et~al.(2013)Iglesias, Konukoglu, Zikic, Glocker, Van~Leemput,
  and Fischl]{iglesias_is_2013}
Juan~Eugenio Iglesias, Ender Konukoglu, Darko Zikic, Ben Glocker, Koen
  Van~Leemput, and Bruce Fischl.
\newblock Is synthesizing {MRI} contrast useful for inter-modality analysis?
\newblock In \emph{International {Conference} on {Medical} {Image} {Computing}
  and {Computer}-{Assisted} {Intervention}}, pages 631--638. Springer, 2013.

\bibitem[Ioffe and Szegedy(2015)]{ioffe_batch_2015}
Sergey Ioffe and Christian Szegedy.
\newblock Batch {Normalization}: {Accelerating} {Deep} {Network} {Training} by
  {Reducing} {Internal} {Covariate} {Shift}.
\newblock \emph{arXiv:1502.03167 [cs]}, March 2015.
\newblock arXiv: 1502.03167.

\bibitem[Jenkinson et~al.(2012)Jenkinson, Beckmann, Behrens, Woolrich, and
  Smith]{jenkinson_fsl_2012}
Mark Jenkinson, Christian Beckmann, Timothy~E. Behrens, Mark Woolrich, and
  Stephen Smith.
\newblock {FSL}.
\newblock \emph{NeuroImage}, 62\penalty0 (2):\penalty0 782--790, August 2012.
\newblock ISSN 1095-9572.
\newblock \doi{10.1016/j.neuroimage.2011.09.015}.

\bibitem[Jog and Fischl(2018)]{jog_pulse_2018}
Amod Jog and Bruce Fischl.
\newblock Pulse {Sequence} {Resilient} {Fast} {Brain} {Segmentation}.
\newblock In \emph{Medical {Image} {Computing} and {Computer} {Assisted}
  {Intervention} – {MICCAI} 2018}, Lecture {Notes} in {Computer} {Science},
  pages 654--662, Cham, 2018. Springer International Publishing.
\newblock ISBN 978-3-030-00931-1.
\newblock \doi{10.1007/978-3-030-00931-1_75}.

\bibitem[Kamnitsas et~al.(2017{\natexlab{a}})Kamnitsas, Baumgartner, Ledig,
  Newcombe, Simpson, Kane, Menon, Nori, Criminisi, Rueckert, and
  Glocker]{kamnitsas_unsupervised_2017}
Konstantinos Kamnitsas, Christian Baumgartner, Christian Ledig, Virginia
  Newcombe, Joanna Simpson, Andrew Kane, David Menon, Aditya Nori, Antonio
  Criminisi, Daniel Rueckert, and Ben Glocker.
\newblock Unsupervised {Domain} {Adaptation} in {Brain} {Lesion} {Segmentation}
  with {Adversarial} {Networks}.
\newblock In \emph{Information {Processing} in {Medical} {Imaging}}, Lecture
  {Notes} in {Computer} {Science}, pages 597--609, Cham, 2017{\natexlab{a}}.
  Springer International Publishing.
\newblock ISBN 978-3-319-59050-9.
\newblock \doi{10.1007/978-3-319-59050-9_47}.

\bibitem[Kamnitsas et~al.(2017{\natexlab{b}})Kamnitsas, Ledig, Newcombe,
  Simpson, Kane, Menon, Rueckert, and Glocker]{kamnitsas_efficient_2017}
Konstantinos Kamnitsas, Christian Ledig, Virginia Newcombe, Joanna Simpson,
  Andrew Kane, David Menon, Daniel Rueckert, and Ben Glocker.
\newblock Efficient multi-scale {3D} {CNN} with fully connected {CRF} for
  accurate brain lesion segmentation.
\newblock \emph{Medical Image Analysis}, 36:\penalty0 61--78, February
  2017{\natexlab{b}}.
\newblock ISSN 1361-8415.
\newblock \doi{10.1016/j.media.2016.10.004}.

\bibitem[Karani et~al.(2018)Karani, Chaitanya, Baumgartner, and
  Konukoglu]{karani_lifelong_2018}
Neerav Karani, Krishna Chaitanya, Christian Baumgartner, and Ender Konukoglu.
\newblock A {Lifelong} {Learning} {Approach} to {Brain} {MR} {Segmentation}
  {Across} {Scanners} and {Protocols}.
\newblock In \emph{Medical {Image} {Computing} and {Computer} {Assisted}
  {Intervention} – {MICCAI} 2018}, Lecture {Notes} in {Computer} {Science},
  pages 476--484, Cham, 2018. Springer International Publishing.
\newblock ISBN 978-3-030-00928-1.
\newblock \doi{10.1007/978-3-030-00928-1_54}.

\bibitem[Krebs et~al.(2019)Krebs, Delingette, Mailhé, Ayache, and
  Mansi]{krebs_learning_2019}
Julian Krebs, Hervé Delingette, Boris Mailhé, Nicholas Ayache, and Tommaso
  Mansi.
\newblock Learning a {Probabilistic} {Model} for {Diffeomorphic}
  {Registration}.
\newblock \emph{IEEE Transactions on Medical Imaging}, 38\penalty0
  (9):\penalty0 2165--2176, September 2019.
\newblock ISSN 0278-0062, 1558-254X.
\newblock \doi{10.1109/TMI.2019.2897112}.

\bibitem[Larsen et~al.(2014)Larsen, Iglesias, and Van~Leemput]{larsen_n3_2014}
Christian Larsen, J.~Eugenio Iglesias, and Koen Van~Leemput.
\newblock N3 {Bias} {Field} {Correction} {Explained} as a {Bayesian} {Modeling}
  {Method}.
\newblock In \emph{Bayesian and {grAphical} {Models} for {Biomedical}
  {Imaging}}, Lecture {Notes} in {Computer} {Science}, pages 1--12, Cham, 2014.
  Springer International Publishing.
\newblock ISBN 978-3-319-12289-2.
\newblock \doi{10.1007/978-3-319-12289-2_1}.

\bibitem[Maes et~al.(1997)Maes, Collignon, Vandermeulen, Marchal, and
  Suetens]{maes_multimodality_1997}
F.~Maes, A.~Collignon, D.~Vandermeulen, G.~Marchal, and P.~Suetens.
\newblock Multimodality image registration by maximization of mutual
  information.
\newblock \emph{IEEE Transactions on Medical Imaging}, 16\penalty0
  (2):\penalty0 187--198, April 1997.
\newblock ISSN 1558-254X.
\newblock \doi{10.1109/42.563664}.

\bibitem[Marcus et~al.(2007)Marcus, Wang, Parker, Csernansky, Morris, and
  Buckner]{marcus_open_2007}
Daniel Marcus, Tracy Wang, Jamie Parker, John Csernansky, John Morris, and
  Randy Buckner.
\newblock Open {Access} {Series} of {Imaging} {Studies} ({OASIS}):
  {Cross}-sectional {MRI} {Data} in {Young}, {Middle} {Aged}, {Nondemented},
  and {Demented} {Older} {Adults}.
\newblock \emph{Journal of cognitive neuroscience}, 19:\penalty0 1498--507,
  October 2007.
\newblock \doi{10.1162/jocn.2007.19.9.1498}.

\bibitem[Marek et~al.(2011)Marek, Jennings, Lasch, Siderowf, Tanner, Simuni,
  Coffey, Kieburtz, Flagg, Chowdhury, Poewe, Mollenhauer, Sherer, Frasier,
  Meunier, Rudolph, Casaceli, Seibyl, Mendick, Schuff, Zhang, Toga, Crawford,
  Ansbach, Blasio, Piovella, Trojanowski, Shaw, Singleton, Hawkins, Eberling,
  Russell, Leary, Factor, Sommerfeld, Hogarth, Pighetti, Williams, Standaert,
  Guthrie, Hauser, Delgado, Jankovic, Hunter, Stern, Tran, Leverenz, Baca,
  Frank, Thomas, Richard, Deeley, Rees, Sprenger, Lang, Shill, Obradov,
  Fernandez, Winters, Berg, Gauss, Galasko, Fontaine, Mari, Gerstenhaber,
  Brooks, Malloy, Barone, Longo, Comery, Ravina, Grachev, Gallagher, Collins,
  Widnell, Ostrowizki, Fontoura, La-Roche, Ho, Luthman, Brug, Reith, and
  Taylor]{marek_parkinson_2011}
Kenneth Marek, Danna Jennings, Shirley Lasch, Andrew Siderowf, Caroline Tanner,
  Tanya Simuni, Chris Coffey, Karl Kieburtz, Emily Flagg, Sohini Chowdhury,
  Werner Poewe, Brit Mollenhauer, Todd Sherer, Mark Frasier, Claire Meunier,
  Alice Rudolph, Cindy Casaceli, John Seibyl, Susan Mendick, Norbert Schuff,
  Ying Zhang, Arthur Toga, Karen Crawford, Alison Ansbach, Pasquale~de Blasio,
  Michele Piovella, John Trojanowski, Les Shaw, Andrew Singleton, Keith
  Hawkins, Jamie Eberling, David Russell, Laura Leary, Stewart Factor, Barbara
  Sommerfeld, Penelope Hogarth, Emily Pighetti, Karen Williams, David
  Standaert, Stephanie Guthrie, Robert Hauser, Holly Delgado, Joseph Jankovic,
  Christine Hunter, Matthew Stern, Baochan Tran, Jim Leverenz, Marne Baca, Sam
  Frank, Cathi~Ann Thomas, Irene Richard, Cheryl Deeley, Linda Rees, Fabienne
  Sprenger, Elisabeth Lang, Holly Shill, Sanja Obradov, Hubert Fernandez,
  Adrienna Winters, Daniela Berg, Katharina Gauss, Douglas Galasko, Deborah
  Fontaine, Zoltan Mari, Melissa Gerstenhaber, David Brooks, Sophie Malloy,
  Paolo Barone, Katia Longo, Tom Comery, Bernard Ravina, Igor Grachev, Kim
  Gallagher, Michelle Collins, Katherine~L. Widnell, Suzanne Ostrowizki, Paulo
  Fontoura, F.~Hoffmann La-Roche, Tony Ho, Johan Luthman, Marcel van~der Brug,
  Alastair~D. Reith, and Peggy Taylor.
\newblock The {Parkinson} {Progression} {Marker} {Initiative} ({PPMI}).
\newblock \emph{Progress in Neurobiology}, 95\penalty0 (4):\penalty0 629--635,
  December 2011.
\newblock ISSN 0301-0082.
\newblock \doi{10.1016/j.pneurobio.2011.09.005}.

\bibitem[Milletari et~al.(2016)Milletari, Navab, and
  Ahmadi]{milletari_v-net_2016}
Fausto Milletari, Nassir Navab, and Seyed-Ahmad Ahmadi.
\newblock V-{Net}: {Fully} {Convolutional} {Neural} {Networks} for {Volumetric}
  {Medical} {Image} {Segmentation}.
\newblock In \emph{2016 {Fourth} {International} {Conference} on {3D} {Vision}
  ({3DV})}, pages 565--571, October 2016.
\newblock \doi{10.1109/3DV.2016.79}.

\bibitem[Moler and Van~Loan(2003)]{moler_nineteen_2003}
Cleve. Moler and Charles. Van~Loan.
\newblock Nineteen {Dubious} {Ways} to {Compute} the {Exponential} of a
  {Matrix}, {Twenty}-{Five} {Years} {Later}.
\newblock \emph{SIAM Review}, 45\penalty0 (1):\penalty0 3--49, January 2003.
\newblock ISSN 0036-1445.
\newblock \doi{10.1137/S00361445024180}.

\bibitem[Patenaude et~al.(2011)Patenaude, Smith, Kennedy, and
  Jenkinson]{patenaude_bayesian_2011}
Brian Patenaude, Stephen Smith, David Kennedy, and Mark Jenkinson.
\newblock A {Bayesian} model of shape and appearance for subcortical brain
  segmentation.
\newblock \emph{NeuroImage}, 56\penalty0 (3):\penalty0 907--922, June 2011.
\newblock ISSN 1053-8119.
\newblock \doi{10.1016/j.neuroimage.2011.02.046}.

\bibitem[Puonti et~al.(2016)Puonti, Iglesias, and
  Van~Leemput]{puonti_fast_2016}
Oula Puonti, Juan~Eugenio Iglesias, and Koen Van~Leemput.
\newblock Fast and sequence-adaptive whole-brain segmentation using parametric
  {Bayesian} modeling.
\newblock \emph{NeuroImage}, 143:\penalty0 235--249, December 2016.
\newblock ISSN 1053-8119.
\newblock \doi{10.1016/j.neuroimage.2016.09.011}.

\bibitem[Rohlfing et~al.(2004)Rohlfing, Brandt, Menzel, and
  Maurer]{rohlfing_evaluation_2004}
Torsten Rohlfing, Robert Brandt, Randolf Menzel, and Calvin Maurer.
\newblock Evaluation of atlas selection strategies for atlas-based image
  segmentation with application to confocal microscopy images of bee brains.
\newblock \emph{NeuroImage}, 21\penalty0 (4):\penalty0 1428--1442, April 2004.
\newblock ISSN 1053-8119.
\newblock \doi{10.1016/j.neuroimage.2003.11.010}.

\bibitem[Ronneberger et~al.(2015)Ronneberger, Fischer, and
  Brox]{ronneberger_u-net_2015}
Olaf Ronneberger, Philipp Fischer, and Thomas Brox.
\newblock U-{Net}: {Convolutional} {Networks} for {Biomedical} {Image}
  {Segmentation}.
\newblock In \emph{Medical {Image} {Computing} and {Computer}-{Assisted}
  {Intervention} – {MICCAI} 2015}, Lecture {Notes} in {Computer} {Science},
  pages 234--241, Cham, 2015. Springer International Publishing.
\newblock ISBN 978-3-319-24574-4.
\newblock \doi{10.1007/978-3-319-24574-4_28}.

\bibitem[Roy et~al.(2019)Roy, Conjeti, Navab, Wachinger, Initiative, and
  {others}]{roy_quicknat_2019}
Abhijit~Guha Roy, Sailesh Conjeti, Nassir Navab, Christian Wachinger,
  Alzheimer's Disease~Neuroimaging Initiative, and {others}.
\newblock {QuickNAT}: {A} fully convolutional network for quick and accurate
  segmentation of neuroanatomy.
\newblock \emph{NeuroImage}, 186:\penalty0 713--727, 2019.
\newblock Publisher: Elsevier.

\bibitem[Sabuncu et~al.(2010)Sabuncu, Yeo, Van~Leemput, Fischl, and
  Golland]{sabuncu_generative_2010}
Mert Sabuncu, Thomas Yeo, Koen Van~Leemput, Bruce Fischl, and Polina Golland.
\newblock A {Generative} {Model} for {Image} {Segmentation} {Based} on {Label}
  {Fusion}.
\newblock \emph{IEEE Transactions on Medical Imaging}, 29\penalty0
  (10):\penalty0 1714--1729, October 2010.
\newblock ISSN 1558-254X.
\newblock \doi{10.1109/TMI.2010.2050897}.

\bibitem[Tae et~al.(2008)Tae, Kim, Lee, Nam, and Kim]{tae_validation_2008}
Woo Tae, Sam Kim, Kang Lee, Eui Nam, and Keun Kim.
\newblock Validation of hippocampal volumes measured using a manual method and
  two automated methods ({FreeSurfer} and {IBASPM}) in chronic major depressive
  disorder.
\newblock \emph{Neuroradiology}, 50\penalty0 (7):\penalty0 569--581, July 2008.
\newblock ISSN 1432-1920.

\bibitem[Van~Leemput et~al.(1999)Van~Leemput, Maes, Vandermeulen, and
  Suetens]{van_leemput_automated_1999}
K.~Van~Leemput, F.~Maes, D.~Vandermeulen, and P.~Suetens.
\newblock Automated model-based tissue classification of {MR} images of the
  brain.
\newblock \emph{IEEE Transactions on Medical Imaging}, 18\penalty0
  (10):\penalty0 897--908, October 1999.
\newblock ISSN 1558-254X.
\newblock \doi{10.1109/42.811270}.

\bibitem[Wells et~al.(1996)Wells, Grimson, Kikinis, and
  Jolesz]{wells_adaptive_1996}
W.M. Wells, W.~Grimson, R.~Kikinis, and F.~Jolesz.
\newblock Adaptive segmentation of {MRI} data.
\newblock \emph{IEEE Transactions on Medical Imaging}, 15\penalty0
  (4):\penalty0 429--442, August 1996.
\newblock ISSN 1558-254X.
\newblock \doi{10.1109/42.511747}.

\bibitem[Zhang et~al.(2001)Zhang, Brady, and Smith]{zhang_segmentation_2001}
Y.~Zhang, M.~Brady, and S.~Smith.
\newblock Segmentation of brain {MR} images through a hidden {Markov} random
  field model and the expectation-maximization algorithm.
\newblock \emph{IEEE transactions on medical imaging}, 20\penalty0
  (1):\penalty0 45--57, January 2001.
\newblock ISSN 0278-0062.
\newblock \doi{10.1109/42.906424}.

\bibitem[Zhao et~al.(2019)Zhao, Balakrishnan, Durand, Guttag, and
  Dalca]{zhao_data_2019}
Amy Zhao, Guha Balakrishnan, Fredo Durand, John Guttag, and Adrian Dalca.
\newblock Data {Augmentation} {Using} {Learned} {Transformations} for
  {One}-{Shot} {Medical} {Image} {Segmentation}.
\newblock In \emph{Proceedings of the {IEEE} {Conference} on {Computer}
  {Vision} and {Pattern} {Recognition}}, pages 8543--8553, 2019.

\bibitem[Çiçek et~al.(2016)Çiçek, Abdulkadir, Lienkamp, Brox, and
  Ronneberger]{cicek_3d_2016}
Özgün Çiçek, Ahmed Abdulkadir, Soeren Lienkamp, Thomas Brox, and Olaf
  Ronneberger.
\newblock {3D} {U}-{Net}: {Learning} {Dense} {Volumetric} {Segmentation} from
  {Sparse} {Annotation}.
\newblock In \emph{Medical {Image} {Computing} and {Computer}-{Assisted}
  {Intervention} – {MICCAI} 2016}, Lecture {Notes} in {Computer} {Science},
  pages 424--432. Springer International Publishing, 2016.
\newblock ISBN 978-3-319-46723-8.

\end{thebibliography}
\end{document}